\documentclass[conference]{IEEEtran}
\IEEEoverridecommandlockouts

\usepackage{cite}
\usepackage{amsmath,amssymb}
\usepackage{graphicx}
\usepackage{booktabs}
\usepackage{array}
\usepackage{url}
\usepackage{tikz}
\usetikzlibrary{arrows.meta,positioning,fit,calc}
\usepackage{pgfplots}
\pgfplotsset{compat=1.17}
\usepackage[hidelinks]{hyperref}

\begin{document}

\title{Daedalus-150M: A Convolution--Attention Hybrid\\
Designed for CPU Inference}

\author{\IEEEauthorblockN{Christos Koutsiaris}
\IEEEauthorblockA{\textit{Development Expert}}
}

\maketitle

\begin{abstract}
Small language models are usually built like large ones and then squeezed onto a
CPU afterwards. We did the opposite: we fixed the target first --- one user, one
token at a time, 4-bit weights, ordinary CPU --- and chose the architecture to
suit it. The result keeps full attention in only 6 of its 18 blocks. The other
12 use short convolutions whose memory is two timesteps wide no matter how long
the conversation gets, so two thirds of the network never re-reads a growing
cache.

Trained from scratch on 59.9\,B tokens, the model scores \textbf{47.31} on a
five-task benchmark against a bar of 42.20 that was fixed before training began.
It beats GPT-2 124M, Pythia-160M, OPT-125M and GPT-neo-125M, all trained on
three to six times more data, and exceeds MobileLLM-125M's published score
despite that model seeing a trillion tokens. Validation bits-per-byte is 0.8685.

To check the architecture rather than the training recipe, we trained a
conventional all-attention model of the same size on the same data, and wrote
down the winning condition before scoring either. The hybrid won the chosen
quality metric by 0.81\,\%, matched it on downstream tasks, produced a 6.3\,\%
smaller 4-bit file, and decoded \textbf{1.76$\times$ faster} at 2048 tokens of
context --- 2.08$\times$ against an external model of similar size. In every
measurement the speed advantage is near zero at an empty context and grows with
length, which is what the mechanism predicts and what a merely leaner model
would not show. A simple bandwidth calculation predicts only 1.17$\times$, so
memory volume alone does not explain the gap.

We also report what did not work: an unmitigated 4-bit quality cost, roughly
half the convolution channels ending up inert and impossible to remove, and a
vocabulary larger than this model size warrants.
\end{abstract}


\section{Introduction}

A language model serving a single user on a CPU is not a smaller version of a
model serving many users on a GPU. Three properties of the regime differ, and
together they change what an efficient architecture looks like.

\textbf{Batch size is one.} There is no batching over which to amortise weight
loading. Every decoded token requires streaming the model's weights through the
memory hierarchy, so throughput is governed by bytes read per token rather than
by arithmetic throughput.

\textbf{Memory bandwidth binds before compute.} A contemporary CPU can issue far
more arithmetic than it can feed with operands. This inverts the usual
optimisation target: an architecture performing more arithmetic over fewer bytes
runs faster.

\textbf{The key--value cache is a growing tax.} In an all-attention decoder,
generating each token requires re-reading the keys and values of every preceding
token in every layer. That cost is linear in context length and is paid per
token. Under high-bandwidth, large-batch GPU serving it is tolerable; at batch
size one on a CPU it dominates long-context decoding.

The last property is the design lever this work pulls. If most layers carry a
\emph{constant-size} state rather than a growing one, decode cost becomes far
flatter in context length, and the advantage widens precisely where users
perceive latency: long conversations and long documents.

\subsection{Contributions}

\begin{enumerate}
\item A concrete sub-200M architecture whose layer composition is chosen from a
      CPU memory-traffic argument rather than adapted from GPU practice.
\item A pre-registered, parameter-matched ablation isolating the effect of that
      composition, with the decision rule fixed before either arm was scored.
\item Decode measurements at three context depths against both an internal twin
      and an external peer, showing the advantage grows with depth in both.
\item A first-order cost model that under-predicts the measured advantage,
      localising the residual to latency rather than bandwidth.
\item A negative result closing structural pruning of dead convolution channels,
      and a full account of five deviations forced on the training run.
\end{enumerate}

\subsection{Success criteria fixed in advance}

To prevent post-hoc rationalisation, the evaluation bar was fixed by operator
decision before any headline number existed. Quality is the five-task mean over
HellaSwag, ARC-Easy, PIQA, OpenBookQA and WinoGrande, scored under
lm-evaluation-harness conventions on a single harness so that peers and this
model are measured identically. Table~\ref{tab:peers} lists the comparison set.

\begin{table}[t]
\caption{Peer set and the pre-registered bar. Scores are measured on this
project's harness, not quoted from publications.}
\label{tab:peers}
\centering
\begin{tabular}{@{}lrrl@{}}
\toprule
\textbf{Model} & \textbf{Tokens} & \textbf{5-task} & \textbf{Target} \\
\midrule
Pythia-160M      & 300\,B & 41.0 & beat \\
GPT-neo-125M     & 300\,B & 41.9 & beat \\
OPT-125M         & 180\,B & 42.1 & beat \\
\textbf{GPT-2 124M} & --- & \textbf{42.2} & \textbf{the bar} \\
MobileLLM-125M   & 1\,T   & gated & stretch \\
Peer-135M        & 2\,T   & 51.2 & concede quality \\
\bottomrule
\end{tabular}
\end{table}

The bar is \textbf{42.2}, the strongest of the four beatable peers. Clearing it
means matching or exceeding 300\,B-token-class models on roughly seven times
less data. The second half of the bar---CPU decode speed---is an architectural
property and does not depend on training outcome.

\section{Related Work}

\textbf{Small language models.} A line of work has shown that sub-1\,B models
improve far past classical compute-optimal token budgets when trained on
carefully filtered data. Pythia~\cite{pythia} established reproducible small
baselines; MobileLLM~\cite{mobilellm} showed that depth-over-width and embedding
sharing matter disproportionately below 350\,M parameters, a finding this design
adopts in its narrow feed-forward and tied embeddings.

\textbf{Recurrent and hybrid sequence models.} Structured state-space
models~\cite{mamba} and gated linear-recurrent
architectures~\cite{rwkv} replace attention with mechanisms whose inference state
is constant in sequence length. Purely recurrent models, however, lose the
precise associative recall that attention provides. Hybrids that retain a
minority of attention layers---Griffin~\cite{griffin} being a prominent
example---recover most of that capability while keeping the majority of layers
cache-free. Daedalus sits in this family, with two distinguishing choices: the
recurrent operator is a very short depthwise convolution rather than a
selective-scan state-space layer, chosen because it maps onto existing CPU
inference kernels without new operators; and the attention-to-recurrence ratio
is fixed by a memory-traffic argument specific to batch-size-one CPU decoding.

\textbf{Efficient attention.} Multi-query and grouped-query
attention~\cite{gqa} shrink the cache by sharing key--value heads across query
heads. This is complementary rather than alternative: Daedalus applies GQA to
the six attention layers it retains, compounding the two reductions.

\textbf{Quantisation.} Post-training quantisation to 4 bits is standard for CPU
deployment. We target the reference runtime's~\cite{llamacpp} \texttt{Q4\_0}
format specifically because its dot-product kernels are the best optimised on
the target hardware, rather than a format with better theoretical
error at lower throughput.

\textbf{What is new here.} The contribution is not a new operator. It is the
observation that in the batch-size-one CPU regime the \emph{ratio} of cache-free
to attention layers is the dominant deployable design variable, together with a
controlled experiment that isolates its effect at matched parameter count.

\section{Architecture}

Daedalus-150M has 160.49\,M parameters in 18 blocks at $d_{\text{model}}=768$,
feed-forward inner dimension 2048, vocabulary 49{,}152 and context 2048. Six
blocks use full attention and twelve use short convolutions, interleaved as

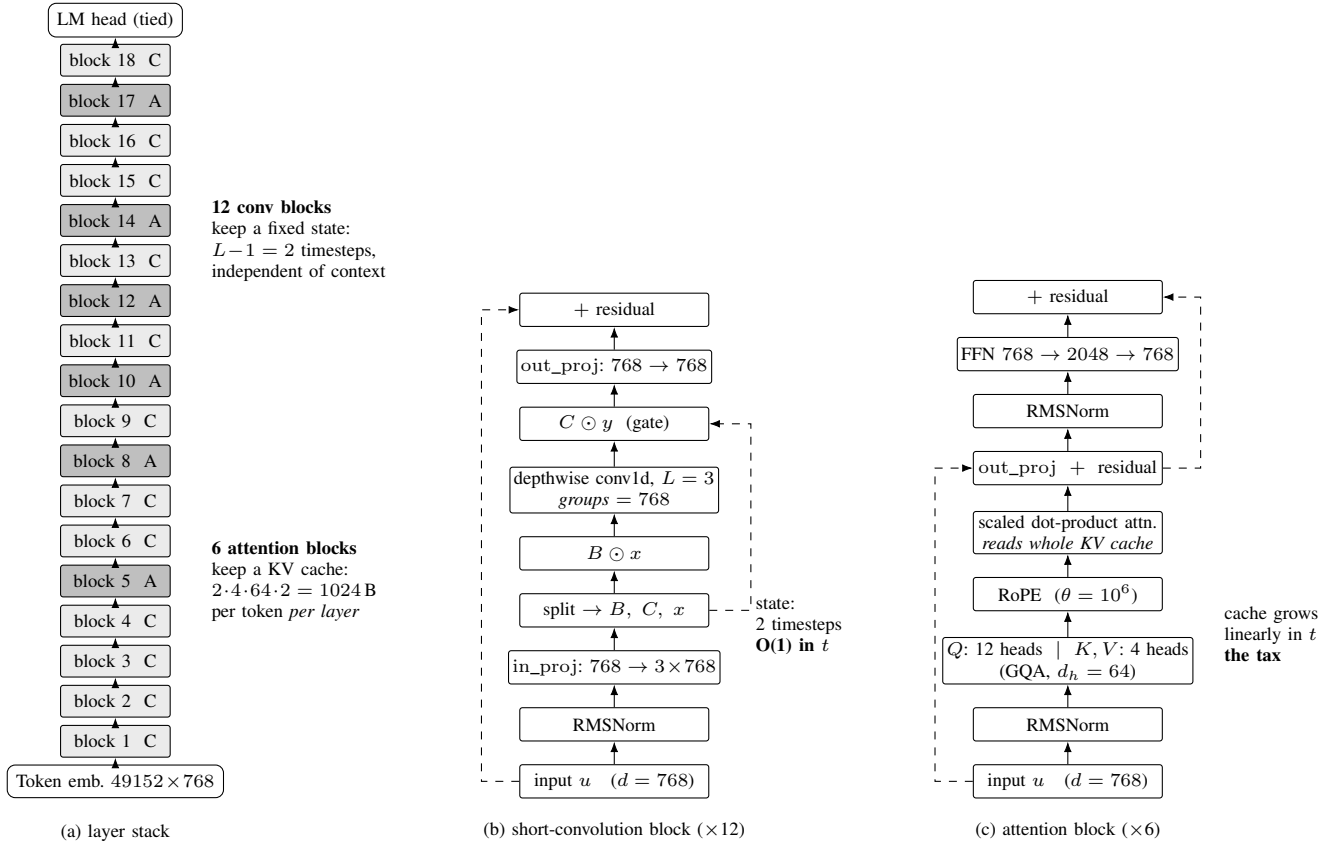
\begin{figure*}[t]
\centering
\begin{tikzpicture}[
  font=\scriptsize,
  blk/.style={draw, rounded corners=1pt, minimum width=1.45cm, minimum height=0.42cm,
              align=center, inner sep=1pt},
  conv/.style={blk, fill=black!8},
  attn/.style={blk, fill=black!25},
  op/.style={draw, rounded corners=1pt, minimum width=2.5cm, minimum height=0.44cm,
             align=center, fill=white, inner sep=1.5pt},
  io/.style={draw, rounded corners=3pt, minimum width=1.45cm, minimum height=0.42cm,
             align=center, fill=white},
  ar/.style={-{Latex[length=1.4mm]}},
]

\node[io] (emb) at (0,0) {Token emb.\ $49152\!\times\!768$};
\def\ystep{0.53}
\foreach \i/\typ/\lab in {
  1/conv/C, 2/conv/C, 3/conv/C, 4/conv/C, 5/attn/A, 6/conv/C,
  7/conv/C, 8/attn/A, 9/conv/C, 10/attn/A, 11/conv/C, 12/attn/A,
  13/conv/C, 14/attn/A, 15/conv/C, 16/conv/C, 17/attn/A, 18/conv/C}
{
  \pgfmathsetmacro{\yy}{\i*\ystep}
  \node[\typ] (b\i) at (0,\yy) {block \i\ \ \lab};
}
\node[io] (head) at (0,{19*\ystep}) {LM head (tied)};
\draw[ar] (emb) -- (b1);
\foreach \i [evaluate=\i as \j using int(\i+1)] in {1,...,17}{\draw[ar] (b\i) -- (b\j);}
\draw[ar] (b18) -- (head);

\node[align=left, anchor=west] at (1.15,{5*\ystep})
  {\textbf{6 attention blocks}\\keep a KV cache:\\$2\!\cdot\!4\!\cdot\!64\!\cdot\!2 = 1024$\,B\\per token \emph{per layer}};
\node[align=left, anchor=west] at (1.15,{13.5*\ystep})
  {\textbf{12 conv blocks}\\keep a fixed state:\\$L\!-\!1 = 2$ timesteps,\\independent of context};
\node[anchor=north] at (0,-0.45) {(a) layer stack};

\begin{scope}[xshift=6.6cm]
\node[op] (c0) at (0,0) {input $u$ \ \ ($d=768$)};
\node[op, above=0.30 of c0] (c1) {RMSNorm};
\node[op, above=0.30 of c1] (c2) {$\operatorname{in\_proj}$: $768 \to 3\!\times\!768$};
\node[op, above=0.30 of c2] (c3) {split $\to B,\;C,\;x$};
\node[op, above=0.30 of c3] (c4) {$B \odot x$};
\node[op, above=0.30 of c4] (c5) {depthwise conv1d, $L=3$\\ \textit{groups} $=768$};
\node[op, above=0.34 of c5] (c6) {$C \odot y$ \ (gate)};
\node[op, above=0.30 of c6] (c7) {$\operatorname{out\_proj}$: $768 \to 768$};
\node[op, above=0.30 of c7] (c8) {$+$ residual};
\foreach \a/\b in {c0/c1,c1/c2,c2/c3,c3/c4,c4/c5,c5/c6,c6/c7,c7/c8}{\draw[ar] (\a) -- (\b);}
\draw[ar, dashed] (c3.east) -- ++(0.55,0) |- (c6.east);
\draw[ar, dashed] (c0.west) -- ++(-0.5,0) |- (c8.west);
\node[anchor=north] at (0,-0.42) {(b) short-convolution block ($\times$12)};
\node[align=left, anchor=west, font=\scriptsize] at (1.75,{2.05}) {state:\\2 timesteps\\\textbf{O(1) in $t$}};
\end{scope}

\begin{scope}[xshift=12.6cm]
\node[op] (a0) at (0,0) {input $u$ \ \ ($d=768$)};
\node[op, above=0.30 of a0] (a1) {RMSNorm};
\node[op, above=0.30 of a1] (a2) {$Q$: 12 heads $\;|\;$ $K,V$: 4 heads\\(GQA, $d_h = 64$)};
\node[op, above=0.34 of a2] (a3) {RoPE \ ($\theta = 10^{6}$)};
\node[op, above=0.30 of a3] (a4) {scaled dot-product attn.\\\textit{reads whole KV cache}};
\node[op, above=0.34 of a4] (a5) {$\operatorname{out\_proj}$ $\;+\;$ residual};
\node[op, above=0.30 of a5] (a6) {RMSNorm};
\node[op, above=0.30 of a6] (a7) {FFN $768 \to 2048 \to 768$};
\node[op, above=0.30 of a7] (a8) {$+$ residual};
\foreach \a/\b in {a0/a1,a1/a2,a2/a3,a3/a4,a4/a5,a5/a6,a6/a7,a7/a8}{\draw[ar] (\a) -- (\b);}
\draw[ar, dashed] (a0.west) -- ++(-0.5,0) |- (a5.west);
\draw[ar, dashed] (a5.east) -- ++(0.5,0) |- (a8.east);
\node[anchor=north] at (0,-0.42) {(c) attention block ($\times$6)};
\node[align=left, anchor=west, font=\scriptsize] at (1.95,{1.95}) {cache grows\\linearly in $t$\\\textbf{the tax}};
\end{scope}

\end{tikzpicture}
\caption{Daedalus-150M. (a) The 18-block stack: 12 short-convolution blocks
(light) interleaved with 6 attention blocks (dark) at indices 4, 7, 9, 11, 13
and 16, between a tied embedding and language-model head. (b) A convolution
block carries a two-timestep state whose size is independent of context, so its
decode cost is constant in $t$. (c) An attention block must re-read the whole
key--value cache per generated token, so its cost is linear in $t$; grouped-query
attention (4 KV heads for 12 query heads) reduces the constant but not the
growth. Every block also carries a feed-forward sub-layer; it is drawn once in
(c) for space. The design question this paper answers empirically is what ratio
of (b) to (c) is right when the deployment target is batch-size-one CPU
decoding.}
\label{fig:arch}
\end{figure*}

\begin{center}
\texttt{C C C C A C C A C A C A C A C C A C}
\end{center}

\noindent where \texttt{A} denotes attention (indices 4, 7, 9, 11, 13, 16) and
\texttt{C} denotes convolution. Fig.~\ref{fig:arch} shows the stack and the
internals of both block types.

\subsection{The short-convolution block}

Each convolution block computes
\begin{align}
B, C, x &= \operatorname{in\_proj}(u) \quad \text{(split three ways)} \\
y &= \operatorname{depthwise\_conv1d}(B \odot x) \\
\text{out} &= \operatorname{out\_proj}(C \odot y)
\end{align}
with kernel length $L=3$ and \texttt{groups} equal to the channel count, so the
convolution is depthwise and channels evolve independently. Its recurrent state
is exactly $L-1=2$ timesteps wide regardless of context length. The gating terms
$B$ and $C$ supply the input-dependent behaviour a fixed kernel alone lacks.

This is the crux of the design: \emph{decoding through a convolution block costs
the same at token 2000 as at token 2, whereas decoding through an attention
block does not.}

\subsection{Why six attention layers}

Pure convolution forfeits precise long-range retrieval, the operation attention
performs uniquely well. The design retains six full-attention blocks, spread
through the depth rather than clustered, so retrieval capacity is distributed
across levels of representation.

The ratio is the whole trade. Against a 24-layer all-attention twin, the hybrid
maintains a key--value cache in six layers rather than twenty-four.

\subsection{Supporting choices}

\textbf{Grouped-query attention}~\cite{gqa} with four key--value heads for twelve
query heads reduces the cache on the six attention layers by a further factor of
three. GQA is usually motivated by GPU memory capacity; here the motivation is
bytes per token.

\textbf{Tied embeddings.} At $49{,}152 \times 768$ the embedding matrix holds
37.7\,M parameters, 23\,\% of the model. Tying input and output projections
removes an entire copy from both the artefact and memory traffic.

\textbf{Feed-forward inner dimension 2048} ($2.67\times d_{\text{model}}$ rather
than the conventional $4\times$) shifts parameters into depth and attention and
away from the widest, most bandwidth-hungry tensors.

\subsection{Rejected alternatives}

Mixture-of-experts routing was rejected at design time: no sub-1\,B precedent
exists at this scale, and the reference runtime's mixture path hardcodes a
specific vendor's gating, so the model would not export. Positional-encoding-free
attention was rejected because it breaks the export format. Distillation from a
larger teacher was cancelled on budget grounds---storing teacher logits required
approximately 288\,GB against a 250\,GB disk.

\section{A Cost Model for CPU Decoding}

Before presenting measurements we state what the architecture \emph{should} buy,
so that the measurement can disagree with it.

At batch size one, generating one token requires reading every weight once, plus
the key--value cache of every attention layer over the current context. Let
$W$ be the quantised weight bytes, $L_A$ the number of attention layers,
$h_{kv}$ the key--value heads, $d_h$ the head dimension, $b$ the bytes per cache
element and $t$ the context depth. Bytes read per generated token are
\begin{equation}
M(t) \;=\; W \;+\; \underbrace{2\,L_A\,h_{kv}\,d_h\,b}_{\textstyle \kappa}\;t
\label{eq:cost}
\end{equation}
where $\kappa$ is the cache bytes per context token.

For the hybrid, $L_A=6$, $h_{kv}=4$, $d_h=64$, $b=2$, giving
$\kappa_{\text{hyb}} = 6144$\,B. For the dense twin ($L_A=24$, $h_{kv}=2$,
$d_h=64$), $\kappa_{\text{dense}} = 12{,}288$\,B --- \textbf{exactly twice},
despite the twin's narrower per-layer cache, because it has four times as many
attention layers. Table~\ref{tab:model} evaluates
Eq.~\eqref{eq:cost} against measurement.

\begin{table}[t]
\caption{First-order bandwidth model versus measurement. The model captures the
direction and the growth but substantially under-predicts the magnitude.}
\label{tab:model}
\centering
\begin{tabular}{@{}lrrrr@{}}
\toprule
\textbf{Depth} & \textbf{$M$ hybrid} & \textbf{$M$ dense} & \textbf{Predicted} & \textbf{Measured} \\
\midrule
0    & 100.2\,MB & 106.6\,MB & 1.06$\times$ & 1.20$\times$ \\
512  & 103.3\,MB & 112.8\,MB & 1.09$\times$ & 1.45$\times$ \\
2048 & 112.8\,MB & 131.7\,MB & 1.17$\times$ & \textbf{1.76}$\times$ \\
\bottomrule
\end{tabular}
\end{table}

\textbf{The model is directionally right and quantitatively wrong}, and the
discrepancy widens with depth. Pure byte accounting predicts a 17\,\% advantage
at depth 2048; the measured advantage is 76\,\%. We therefore reject the
hypothesis that bandwidth alone explains the result and attribute the residual to
two effects that Eq.~\eqref{eq:cost} does not model. First, attention traverses
the cache with a dependent softmax reduction, which is latency-bound rather than
bandwidth-bound and scales poorly when the working set exceeds last-level cache;
a depthwise convolution instead streams a two-element state with perfect
locality. Second, the twin executes 24 layers to the hybrid's 18, so per-layer
fixed costs are paid a third more often.

The practical consequence is that the architecture's benefit is
\emph{larger} than a naive byte count suggests, and that improving attention
kernels would narrow but not close the gap---the cache the hybrid does not keep
cannot be optimised.

\section{Corpus and Data Pipeline}

Training data is a ten-source English mixture totalling \textbf{16.93\,B unique
tokens}, weighted toward educational and reasoning-dense text
(Table~\ref{tab:corpus}). At this scale data quality dominates quantity, and the
evaluation suite is knowledge- and commonsense-heavy.

\begin{table}[t]
\caption{Corpus composition (blueprint shares).}
\label{tab:corpus}
\centering
\begin{tabular}{@{}lrl@{}}
\toprule
\textbf{Source} & \textbf{Share} & \textbf{Role} \\
\midrule
FineWeb-Edu            & 37.5\,\% & educational web, backbone \\
DCLM-baseline          & 22.5\,\% & broad filtered web \\
Stack-Edu (Python)     & 9.0\,\%  & code \\
FinePDFs-Edu           & 8.0\,\%  & long-form document prose \\
FinePhrase             & 7.0\,\%  & phrase-level diversity \\
Cosmopedia-v2          & 5.0\,\%  & synthetic textbook style \\
FineMath-3+            & 3.0\,\%  & mathematics \\
InfiWebMath-3+         & 3.0\,\%  & web mathematics \\
FineWiki-en            & 3.0\,\%  & encyclopedic reference \\
Everyday-conversations & 2.0\,\%  & dialogue register \\
\bottomrule
\end{tabular}
\end{table}

\subsection{Bounded repetition}

A 59.9\,B-token budget over a 16.93\,B-token corpus implies roughly 3.5 epochs.
Rather than allow any source to be repeated arbitrarily to meet its nominal
share, each source is capped at four epochs, following the finding that
repetition up to approximately four epochs costs little relative to unique
data~\cite{muennighoff}. Mass freed by a capped source is redistributed by
water-filling across sources retaining headroom.

The cap is load-bearing. Without it, a source such as
\emph{everyday-conversations} (approximately 400\,k tokens) would be repeated
thousands of times to fill a 2\,\% share, which is not what a 2\,\% share is
meant to express.

\subsection{Packing and held-out data}

Documents are tokenised and packed into fixed-size shards of contiguous token
ids. Each source reserves whole shard \emph{files}---never token slices---for a
$\approx$2\,\% holdout, so no training window can straddle the train/holdout
boundary. Because shard files are reserved whole, the realised holdout fraction
varies per source: a source whose trailing shard is small relative to the target
must reserve the preceding shard too, and one source's real carve is 9.16\,\%
rather than 2\,\%. Uneven carves perturb the training mixture, so the effect is
measured rather than assumed.

Validation bits-per-byte is weighted by the probabilities the sampler actually
draws with, not by holdout token counts. Correcting this distinction changed the
reported metric by 9.7\,\%: weighting by holdout size systematically
over-weighted the hardest sources, because those sources happened to have larger
trailing shards.

\section{Training}

Table~\ref{tab:training} summarises the configuration. Parameters are split by
tensor shape: two-dimensional weight matrices are optimised by Muon~\cite{muon},
and embeddings, norms and biases by AdamW---122.68\,M parameters over 102
tensors and 37.81\,M over 62 respectively.

\begin{table}[t]
\caption{Training configuration.}
\label{tab:training}
\centering
\begin{tabular}{@{}ll@{}}
\toprule
Budget      & 59.9\,B tokens, 124{,}476 steps \\
Optimisers  & Muon (122.68\,M params) + AdamW (37.81\,M) \\
Learning rates & Muon 0.02, AdamW $3\times10^{-4}$ \\
Schedule    & WSD, 300-step warmup, linear decay to zero \\
            & over the final 45\,\% \\
Batch ramp  & 128\,k $\rightarrow$ 512\,k tokens/step (first 10\,\%) \\
Sequence ramp & 1024 $\rightarrow$ 2048 (first 10\,\%) \\
Regularisation & $z$-loss $10^{-4}$, gradient clip 1.0 \\
Precision   & bf16 \\
Hardware    & 1$\times$ RTX 5090 (32\,GB) \\
\bottomrule
\end{tabular}
\end{table}

\subsection{Decaying the learning rate to zero}

The schedule is warmup--stable--decay with \emph{linear decay to zero}, not
cosine decay to a floor, following evidence that full decay to zero is
substantially more sample-efficient~\cite{d2z}.

A practical consequence deserves emphasis for anyone reading the loss curve:
\textbf{training loss is expected to plateau during the stable phase.} Between
approximately step 20{,}000 and step 68{,}461 the learning rate is held at its
peak and loss moves very little. This is the schedule operating as designed, not
a stall; nearly all remaining quality is purchased during the decay phase. In
this run the plateau was mistaken for a stall by an observer, which is precisely
why it is documented here.

A secondary benefit is that the checkpoint at the onset of decay is a reusable
branch point: it can be trained further on additional or different data and
re-decayed, whereas a model already annealed to $\text{lr} \approx 0$ requires
learning-rate re-warmup.

\section{Central Experiment: Hybrid versus Dense}

The architectural claim was tested directly, with the decision rule fixed before
either arm was scored. Two models, parameter-matched to within 0.5\,\%
(160.49\,M hybrid against 161.25\,M dense, the twin being 24 all-attention
layers at $d=640$ with feed-forward 2304), were trained on identical data and
schedule for 5\,B tokens each and fully decayed. The pre-registered metric was
validation bits-per-byte over a 645\,M-token held-out set with a 0.5\,\% margin
floor fixed in advance. A dense win beyond that floor was a live possible
outcome that would have changed the main run's architecture.

\subsection{Quality}

\begin{table}[t]
\caption{Ablation quality. The two metrics disagree in direction.}
\label{tab:quality}
\centering
\begin{tabular}{@{}lrr@{}}
\toprule
& \textbf{val\_bpb} $\downarrow$ & \textbf{5-task} \\
\midrule
Daedalus-150M (hybrid) & \textbf{0.910398} & 44.68 \\
dense-150m (twin)      & 0.917774 & \textbf{44.82} \\
\bottomrule
\end{tabular}
\end{table}

The hybrid wins the pre-registered metric by 0.81\,\%, clearing the 0.5\,\%
floor. On the five-task mean the dense twin is nominally ahead by 0.14 points,
approximately $0.24\sigma$ against the $\approx 0.58\sigma$ these suites carry.
Per task the two swap places in both directions---the hybrid leads on HellaSwag
and OpenBookQA, the twin on WinoGrande, PIQA and ARC-Easy---which is the
signature of noise rather than of a pattern.

At a 5\,B-token budget the downstream suite sits near its noise floor:
WinoGrande scores 50.0 and 51.6 against a 50.0 chance floor, so roughly two of
five tasks measure nothing. Validation bits-per-byte, computed over 645\,M
held-out tokens, separates architectures far more sensitively at this scale,
which is why it was pre-registered.

The honest summary is that \textbf{the hybrid matches on quality and wins
decisively on decode}. It is not a downstream-task win, and is not presented as
one.

\subsection{Per-domain difficulty}

Table~\ref{tab:persource} decomposes validation bits-per-byte by source for the
dense arm. The spread is nearly a factor of two and is informative for anyone
reusing this corpus: code and encyclopedic text are far more predictable than
broad filtered web text, so an aggregate figure is dominated by the mixture
weights as much as by the model.

\begin{table}[t]
\caption{Per-source validation bits-per-byte (dense arm, 5\,B tokens).}
\label{tab:persource}
\centering
\begin{tabular}{@{}lrr@{}}
\toprule
\textbf{Source} & \textbf{val\_bpb} & \textbf{Weight} \\
\midrule
Stack-Edu (Python)     & 0.5811 & 9.2\,\% \\
Cosmopedia-v2          & 0.6109 & 5.1\,\% \\
FineWiki-en            & 0.7625 & 3.1\,\% \\
FinePDFs-Edu           & 0.9241 & 8.2\,\% \\
InfiWebMath-3+         & 0.9361 & 3.1\,\% \\
FinePhrase             & 0.9381 & 7.1\,\% \\
FineWeb-Edu            & 0.9451 & 38.3\,\% \\
FineMath-3+            & 0.9662 & 3.1\,\% \\
DCLM-baseline          & 1.0783 & 23.0\,\% \\
\bottomrule
\end{tabular}
\end{table}

\subsection{CPU decode speed}
\label{sec:decode}

Decode is measured with 4-bit weights on 8 threads, generating 128 tokens after
priming a context of the stated depth. Arms alternate within a single pass so
that background load on the measurement machine perturbs both equally; the
ratio within a pass is therefore more trustworthy than the absolute
throughputs, which are depressed by whatever else is running.

\begin{table}[t]
\caption{CPU decode, trained weights, alternating passes. The advantage grows
monotonically with context.}
\label{tab:decode}
\centering
\begin{tabular}{@{}lrrr@{}}
\toprule
\textbf{Depth} & \textbf{Hybrid} & \textbf{Dense twin} & \textbf{Ratio} \\
\midrule
0 (empty)    & $1111.9 \pm 25.7$ & $922.8 \pm 14.2$ & 1.20$\times$ \\
512          & $960.3 \pm 10.7$  & $664.4 \pm 6.5$  & 1.45$\times$ \\
\textbf{2048} & $\mathbf{739.3 \pm 35.5}$ & $\mathbf{420.3 \pm 5.5}$ & \textbf{1.76$\times$} \\
\bottomrule
\end{tabular}
\end{table}

Table~\ref{tab:decode} contains the central result, and its \emph{shape} is the
thesis. At depth zero the hybrid has almost nothing to gain, because its
advantage \emph{is} the cache it does not keep and an empty context has no cache
to re-read. The advantage grows monotonically with context, reaching
1.76$\times$ at the trained context length.

One confound runs against the hybrid: the dense twin receives the reference
runtime's better-optimised attention graph, which has had far more engineering
attention than the hybrid's convolution path. The margin is achieved in spite of
that, and would be expected to widen if the convolution kernels received
comparable optimisation.

Two figures appearing in this project's history should not be quoted:
1.29$\times$ originated from a non-alternating measurement that does not
reproduce, and 1.15--1.17$\times$ is the depth-zero row, a floor rather than a
result.

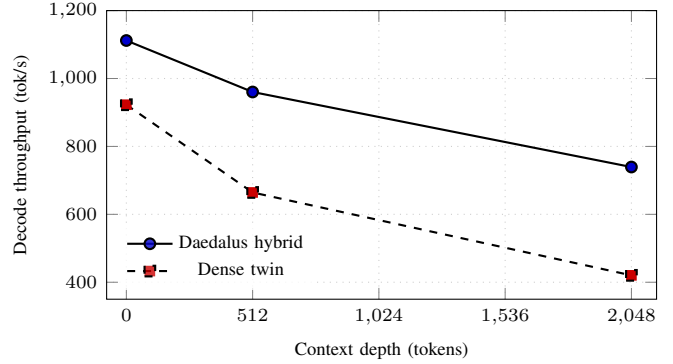
\begin{figure}[t]
\centering
\begin{tikzpicture}
\begin{axis}[
    width=\columnwidth, height=5.4cm,
    xlabel={Context depth (tokens)},
    ylabel={Decode throughput (tok/s)},
    xmin=-80, xmax=2150, ymin=350, ymax=1200,
    xtick={0,512,1024,1536,2048},
    legend style={at={(0.02,0.03)}, anchor=south west, font=\scriptsize,
                  draw=none, fill=white, fill opacity=0.8, text opacity=1},
    grid=major, grid style={dotted, gray!40},
    tick label style={font=\scriptsize},
    label style={font=\scriptsize},
]
\addplot+[mark=*, thick, color=black]
  coordinates {(0,1111.9) (512,960.3) (2048,739.3)};
\addlegendentry{Daedalus hybrid}
\addplot+[mark=square*, thick, color=black, dashed]
  coordinates {(0,922.8) (512,664.4) (2048,420.3)};
\addlegendentry{Dense twin}
\end{axis}
\end{tikzpicture}
\caption{Decode throughput against context depth (4-bit weights, 8 threads,
trained weights). Both curves fall as the context grows, but the dense twin
falls roughly twice as fast: it re-reads $\kappa=12{,}288$\,B of cache per
context token against the hybrid's $6{,}144$\,B. The vertical gap, not the
absolute height, is the architectural result.}
\label{fig:decode}
\end{figure}

Fig.~\ref{fig:decode} plots the same data. The two curves do not merely differ
by a constant: they diverge, and the divergence is the quantity the architecture
was designed to produce.

\subsection{Comparison against an external peer}

\begin{table}[t]
\caption{CPU decode against an external 135\,M-parameter peer, same harness.}
\label{tab:peerdecode}
\centering
\begin{tabular}{@{}lrrr@{}}
\toprule
\textbf{Depth} & \textbf{Daedalus} & \textbf{Peer-135M} & \textbf{Ratio} \\
\midrule
0            & $960.9 \pm 3.4$  & $908.1 \pm 37.1$ & 1.06$\times$ \\
512          & $933.7 \pm 28.5$ & $625.7 \pm 38.5$ & 1.49$\times$ \\
\textbf{2048} & $\mathbf{648.6 \pm 12.6}$ & $\mathbf{312.4 \pm 7.2}$ & \textbf{2.08$\times$} \\
\bottomrule
\end{tabular}
\end{table}

Daedalus decodes 2.08$\times$ faster at the context it is built for while
carrying 19\,\% more parameters. The same signature---near unity at depth zero,
growing with context---reproduces against a different architecture written by a
different author, which is stronger evidence for the mechanism than the
within-project comparison alone, since it cannot be an artefact of this
project's training or export code.

\subsection{Quantised artefact size}

Four-bit output size depends on tensor shapes rather than weight values, so this
result was final before training completed. Matched within 0.5\,\% at half
precision (306.22\,MiB against 307.63\,MiB), the hybrid's shipped file is
95.56\,MiB against the twin's 101.62\,MiB---\textbf{6.3\,\% smaller}, at 4.99
versus 5.29 bits per weight. The quantiser reports that one of 266 tensors
required fallback quantisation on the dense twin and issues no such warning on
the hybrid: a shape effect that happens to favour the hybrid's dimensions, and a
reminder that the deployed artefact size is not a pure function of parameter
count.

\subsection{One datapoint against the bar}

The hybrid ablation arm---identical architecture, fully decayed, but trained on
only 5\,B tokens---scores \textbf{44.7} on the five-task harness. That clears the
42.2 bar and exceeds all four 300\,B-token-class peers on every one of the five
tasks. This is not the Daedalus result and nothing is projected from it; it moves
the open question from whether the architecture can reach the bar to how far
beyond it twelve times more data carries.

\section{Headline Results}

Table~\ref{tab:final} gives the finished model measured on the harness described
in Section~\ref{sec:evalmethod}, alongside the peers it was required to beat.

\begin{table}[t]
\caption{Five-task mean. Every peer was re-scored here rather than quoted from
its own paper, so the column is internally comparable.}
\label{tab:final}
\centering
\begin{tabular}{@{}lrr@{}}
\toprule
\textbf{Model} & \textbf{Training tokens} & \textbf{5-task} \\
\midrule
\textbf{Daedalus-150M} & \textbf{59.9\,B} & \textbf{47.31} \\
MobileLLM-125M   & 1\,T   & 46.3 (published) \\
GPT-2 124M       & ---    & 42.2 \\
OPT-125M         & 180\,B & 42.1 \\
GPT-neo-125M     & 300\,B & 41.9 \\
Pythia-160M      & 300\,B & 41.0 \\
\midrule
Peer-135M        & 2\,T   & 51.2 \\
\bottomrule
\end{tabular}
\end{table}

The model clears the 42.20 bar by 5.11 points and beats every model it was set
against, each of which saw three to six times more data. It also exceeds
MobileLLM-125M's published figure, which came from a trillion tokens.

The 2\,T-token peer in the last row remains ahead by 3.9 points. That was
conceded before training started: this project trades quality against decode
speed at a fixed size, and Section~\ref{sec:decode} is the other half of that
trade.

\begin{table}[t]
\caption{Per-task scores. WinoGrande sits at its chance floor and is measuring
nothing at this scale.}
\label{tab:pertask}
\centering
\begin{tabular}{@{}lrr@{}}
\toprule
\textbf{Task} & \textbf{Score} & \textbf{Chance} \\
\midrule
PIQA        & 65.78 & 50.0 \\
ARC-Easy    & 50.42 & 25.0 \\
WinoGrande  & 50.04 & 50.0 \\
HellaSwag   & 37.93 & 25.0 \\
OpenBookQA  & 32.40 & 25.0 \\
\midrule
\textbf{Mean} & \textbf{47.31} & \\
\bottomrule
\end{tabular}
\end{table}

Validation bits-per-byte over a 645\,M-token held-out set is \textbf{0.8685}.
The same architecture trained on 5\,B tokens reached 0.9104, so the remaining
55\,B tokens bought a 4.6\,\% improvement --- a large gain by the standards of
this metric, which moves slowly.

Quantising to 4 bits costs roughly 6\,\% perplexity (9.18 at half precision
against 9.75 at \texttt{Q4\_0}, measured on held-out encyclopedic text). That is
higher than the 2.5\,\% measured at 5\,B tokens and is discussed in
Section~\ref{sec:qat}.

\section{Limitations}

\subsection{The training mixture drifted from its target}

The corpus holds 16.9\,B unique tokens and the run consumed 59.9\,B, so most
text is seen about three and a half times. Each source is capped at four
repetitions, and for the largest sources that cap binds --- their share cannot
be met without exceeding it, so the freed share flows to smaller sources that
still have unread text.

The realised mixture therefore sits some distance from the intended one.
Measured as $L_1$ distance in percentage points, the run as a whole is at
\textbf{10.42} against a limit of 10.0 that was fixed in advance. The largest
single-source deviation is $-0.24$ points, which is negligible against benchmark
noise, but the headline figure is over the limit and is reported as such. If
downstream results ever look weaker than the peer comparison suggests they
should, mixture drift is a live explanation.

\subsection{The last 8\,\% of training used a slightly smaller corpus}

Training was interrupted and resumed from a checkpoint. Three consequences are
permanent properties of the released weights and are stated rather than
smoothed over.

The optimiser state was not carried across: momentum restarted from zero and
re-warmed over a few hundred steps. This is mild --- by that point the learning
rate had already decayed to 0.0035 --- but the run is not one continuous
optimiser trajectory.

The data cursor also reset, so the final 4.8\,B tokens are a fresh pass over the
corpus rather than a continuation. Some documents were seen twice and some not
at all.

Those tokens came from a corpus snapshot 0.42\,B tokens smaller than the one used
earlier, which is what pushed the mixture figure above from 9.94 to 10.42.

\subsection{Quantisation-aware training did not run}
\label{sec:qat}

The plan was to spend the final 5\,\% of training with fake quantisation applied
on the deployment format's exact 4-bit grid, verified against the runtime's own
kernel rather than reimplemented. It produced a non-finite loss on its first
step and was disabled; the run finished without it.

\textbf{The released model is therefore quantised after training rather than
during it}, and carries the full 4-bit penalty --- about 6\,\% perplexity rather
than the 2.5\,\% measured at smaller scale. The half-precision weights are
published, so this is fixable without repeating any training: either by running
the quantisation-aware pass from the finished checkpoint, or by choosing a
format with better error behaviour at the same bit width. We did not diagnose
the failure.

\subsection{The vocabulary is larger than this size warrants}

The tokeniser, and with it a 49{,}152-entry vocabulary, was adopted from an
existing model because an earlier plan required student and teacher to share
one exactly. That plan was dropped for unrelated reasons; the vocabulary stayed.

Scaling laws put the optimum for a 150\,M model nearer 24--32\,k. At 49{,}152 the
embedding table holds 37.7\,M parameters --- \textbf{23\,\% of the model} --- where
a 32\,k vocabulary would hold 24.6\,M, freeing roughly 13\,M parameters for
layers that compute rather than look up. Tied embeddings recover half the waste,
which is probably why it was never revisited. A smaller vocabulary is the first
thing to change in a successor.

We did verify the tokeniser rather than assume it: reading token ids directly
out of the quantised file gave \textbf{zero mismatches} against the reference
tokeniser. A silent mismatch here ships a model that produces fluent nonsense
while every reported metric looks healthy, and nothing else in the pipeline
would catch it.

\subsection{Single seed, English only}

Every number here comes from one seed. The 0.81\,\% ablation margin is not a
confidence interval, and the downstream comparison is reported as a tie partly
because one seed cannot resolve a quarter of a standard deviation. The model is
English-only, and the decode advantage is measured at the 2048-token context it
was trained for, not extrapolated past it.

\section{A Negative Result: Dead Channels Cannot Be Reclaimed}
\label{sec:dead}

Approximately 47.9\,\% of short-convolution channels contribute nothing to the
model's output. This is a stable plateau rather than progressive decay: the
fraction measures 47.928\,\% at step 9{,}896 and 47.993\,\% at step 30{,}041,
and is unchanged over the following 10{,}743 steps. It represents roughly
13.6\,M inert parameters, an 8.5\,\% parameter inefficiency.

Because the mask is stable and the convolution is depthwise, structural pruning
at export appeared straightforward: narrow each block's projections and kernel
while leaving the residual stream at full width, for an estimated 7.7\,MB
saving. \textbf{This was tested and does not work.} The reference runtime creates
all three short-convolution tensors at fixed model width and shape-checks them at
load time; a real artefact narrowed from 768 to 640 is rejected with
\texttt{check\_tensor\_dims: expected 3,768, got 3,640}, while both the
unmodified file and the same file rebuilt at full width through the identical
writer load correctly. That second control is what makes the finding conclusive:
the rejection is the narrowing itself, not an artefact of the rewriting tool.

Reclaiming the parameters would require patching the inference runtime, which
would forfeit stock-binary compatibility---the foundation of the CPU-decode
claim---in exchange for 7.7\,MB. Four-bit quantisation moreover spends four bits
on a zero exactly as on any other weight, so the inert channels cost nothing
beyond their share of the file. The appropriate remedy belongs to the next
model's initialisation and regularisation, not to a retrofit of this one.

\section{Evaluation Methodology}
\label{sec:evalmethod}

Because several claims in this paper are comparative, the instruments deserve
description.

\subsection{Quality}

The five-task suite is scored under lm-evaluation-harness conventions with
length-normalised accuracy where the harness specifies it. Crucially, \emph{peer
models are re-scored on the same harness} rather than compared against published
figures. The gap between the two is not negligible: published eight-task means
for the peer set run 0.5--1.5 points above what the same checkpoints score here,
because the task subset and normalisation differ. Comparing a locally measured
number against a published one would manufacture roughly a point of spurious
advantage.

Validation bits-per-byte is computed over a 645\,M-token held-out set. Bits per
\emph{byte}, rather than per token, is used so that the figure is comparable
across tokenisers---a model with a more efficient tokeniser would otherwise
appear better at equal predictive quality.

\subsection{Decode}

Decode throughput is measured by generating 128 tokens after priming a context
of the stated depth, on 8 threads, with 4-bit weights, repeated and reported as
mean $\pm$ standard deviation. Two protocol details matter. First, arms
\emph{alternate within a single pass} rather than running back to back: an
earlier back-to-back measurement in this project produced a 1.29$\times$ figure
that did not reproduce, because background load drifted between the two halves.
Second, absolute throughputs are depressed by whatever else shares the machine,
so the ratio within a pass is the trustworthy quantity and the absolute numbers
should be treated as a floor.

The measurement includes only generation, not prompt processing. Prompt
processing is compute-bound and parallel over positions, so it is a regime where
attention is not disadvantaged; including it would flatter the hybrid on
workloads that decode little.

\section{Deployment Notes}

The model targets stock inference binaries with no patches, which constrained
several decisions and is worth making explicit for anyone reproducing the
deployment.

\textbf{Quantisation format.} \texttt{Q4\_0} is chosen over formats with better
error characteristics at equal bit width because its dot-product kernels are the
best optimised on the target hardware. The measured quality cost of that choice
is 2.5\,\% perplexity at 5\,B tokens without quantisation-aware training; the
throughput it buys is what the entire design exists to deliver. Formats such as
\texttt{Q4\_K} would reduce the error and give back part of the speed.

\textbf{Memory footprint.} The shipped artefact is 95.56\,MiB. At 2048-token
context the key--value cache adds approximately 12.6\,MB, so a complete
single-user session fits comfortably under 128\,MB---small enough for the model
to be resident alongside an application rather than requiring a dedicated
process boundary.

\textbf{Thread scaling.} All figures here use 8 threads. Because decoding at
batch size one is memory-bound, throughput saturates once threads suffice to
saturate memory bandwidth; adding cores past that point yields little. The
hybrid saturates at a lower thread count than the dense twin, since it moves
fewer bytes per token, which is a secondary benefit on shared machines.

\section{Discussion}

\textbf{When the hybrid wins.} The advantage is a function of context depth. Any
deployment that keeps a long conversation, retrieves documents into the prompt,
or processes files will sit at the right-hand end of Table~\ref{tab:decode}.
Deployments that issue short, independent prompts into an empty context will see
the depth-0 row and little benefit.

\textbf{Extrapolation beyond 2048.} Eq.~\eqref{eq:cost} implies the ratio keeps
growing with $t$, and the measured trend is consistent with that. We decline to
quote a figure beyond the trained context, because the model has not been
trained there and the claim would not be about a usable model.

\textbf{Where the ratio could be improved.} Two thirds of the layers are already
cache-free, so the remaining cache is concentrated in six layers. Reducing
attention further trades against retrieval capability, which this experiment did
not measure; a retrieval-sensitive evaluation is the right instrument for that
question and is future work.

\textbf{Future work.} Diagnosing the quantisation-aware training failure;
addressing dead channels at initialisation rather than export; a depth ablation
($18\times768$ against $24\times640$) which is designed but unrun; multi-seed
replication of the ablation; and a retrieval-sensitive evaluation to bound how
far the attention fraction can fall.

\section{Reproducibility}

Every number in this paper comes from a file in the public repository. The
head-to-head quality, decode and file-size figures come from the comparison's
own results file; the winning condition comes from a document timestamped before
either model was scored, so it could not have been fitted to the outcome; the
five-task and bits-per-byte figures come from the evaluation outputs; and the
pruning result comes from a script that can be re-run against a real model file.

The model weights are published at half precision as well as 4-bit, so the
quantisation figures can be reproduced and improved on without repeating any
training. The corpus is assembled entirely from public datasets.

\section{Conclusion}

The proposition was specific and falsifiable: for one user decoding on a CPU,
replacing two thirds of a transformer's attention layers with fixed-state short
convolutions buys a large speed advantage at no cost in quality.

Both halves held. The parameter-matched comparison gave the hybrid the chosen
quality metric by 0.81\,\% against a margin fixed in advance, a tie on
downstream tasks, a 6.3\,\% smaller 4-bit file, and \textbf{1.76$\times$ faster
decoding} at 2048 tokens --- 2.08$\times$ against an outside model. The advantage
grows with context in every measurement, which is what identifies the cause: a
model that were simply leaner would be faster by a constant.

At full scale the model scores \textbf{47.31} against a bar of 42.20 set before
training, beating every model in its size class trained on three to six times
more data. Bits-per-byte is 0.8685.

What remains open is not the architecture but the engineering around it: a 4-bit
penalty that quantisation-aware training was meant to remove, roughly half the
convolution channels sitting inert because nothing in training discourages that,
and a vocabulary inherited rather than chosen. None of the three requires new
research to fix, and all three are cheaper to address in the next model than to
retrofit into this one.

\end{document}